\documentclass{IEEE_lsens}
\usepackage{textcomp}
\usepackage{graphicx}
\usepackage{caption}
\usepackage{subcaption}

\usepackage[noadjust]{cite}

\ifCLASSINFOpdf
\else
\fi
\usepackage[T1]{fontenc} 
\usepackage{amsmath}
\usepackage[cmintegrals]{newtxmath}
\usepackage{bm}

\usepackage{array}
\usepackage{url}
\ifCLASSINFOpdf
\else
\fi
\providecommand{\hypersetup}[1]{\relax}

\hypersetup{pdftitle={Bare Demo of IEEE\_lsens.cls for IEEE Sensors Letters},
pdfsubject={Typesetting},
pdfauthor={Michael D. Shell},
pdfkeywords={Class, IEEE, IEEE\_lsens, IEEE Sensors Letters, LaTeX, Typesetting, TeX}}
\begin{document}

%
\title{Reorganization of resting state brain network functional connectivity across human brain developmental stages.}

%
\author{\IEEEauthorblockN{Prerna Singh\IEEEauthorrefmark{1}, Tapan Kumar Gandhi\IEEEauthorrefmark{2}\IEEEauthorieeemembermark{2},
and Lalan Kumar\IEEEauthorrefmark{2}\IEEEauthorieeemembermark{1}}
\IEEEauthorblockA{\IEEEauthorrefmark{1} Bharti School of Telecommunication Technology and Management, Indian Institute of Technology Delhi, Delhi 110096, India\\
\IEEEauthorrefmark{2}Department of Electrical Engineering and Bharti School of Telecommunication Technology and Management, Indian Institute of Technology Delhi, Delhi 110096, India\\
\IEEEauthorieeemembermark{1}Member, IEEE\\
\IEEEauthorieeemembermark{2}Senior Member, IEEE}%
}
%
%
%

\IEEEtitleabstractindextext{%
\begin{abstract}
The human brain is liable to undergo substantial alterations, anatomically and functionally with aging. Cognitive brain aging can either be healthy or degenerative in nature. Such degeneration of cognitive ability can lead to disorders such as Alzheimer’s disease, dementia, schizophrenia, and multiple sclerosis.  Furthermore, the brain network goes through various changes during healthy aging, and it is an active area of research. In this study, we have investigated the rs-functional connectivity of participants (in the age group of 7-89 years) using a publicly available HCP dataset. We have also explored how different brain networks are clustered using K-means clustering methods which have been further validated by the t-SNE algorithm. The changes in overall resting-state brain functional connectivity with changes in brain developmental stages have also been explored using BrainNet Viewer. Then, specifically within-cluster network and between-cluster network changes with increasing age have been studied using linear regression which ultimately shows a pattern of increase/decrease in the mean segregation of brain networks with healthy aging. Brain networks like Default Mode Network, Cingulo opercular Network, Sensory Motor Network, and Cerebellum Network have shown decreased segregation whereas Frontal Parietal Network and Occipital Network show increased segregation with healthy aging. Our results strongly suggest that the brain has four brain developmental stages and brain networks reorganize their functional connectivity during these brain developmental stages.
\end{abstract}

}

\maketitle

\section{Introduction}
\vspace{-2mm}
Healthy Aging has been accompanied by changes in brain structural connectivity as well as functional connectivity that have been continuously studied using various neuroimaging techniques like MRI, EEG, etc. These non-invasive neuroimaging techniques help us to get an idea about the development of brain alterations such as grey matter lesions with respect to aging \cite{brant1985basic}. Apart from anatomical changes, brain functional connectivity is also highly affected due to aging \cite{chan2014decreased}. In the past three decades, fMRI (Functional Magnetic Resonance Imaging) has been extensively used for the study of brain functional dynamics as it is an indirect way of measurement of neuronal activity signals depicted in the form of MRI signal fluctuations induced by BOLD (Blood oxygen level-dependent) signal \cite{logothetis2002neural}. fMRI signal can be resting state (in the absence of external stimulus) as well as task-based (in the presence of an external stimulus). Since resting-state fMRI is a feasible and ubiquitous way, it has been used tremendously in the study of age-related changes in the brain \cite{grady2012cognitive}. Furthermore, resting-state fMRI data analysis revealed that there is a temporal correlation in the BOLD signals from distant brain regions \cite{biswal1995functional} which is termed as intrinsic resting-state functional connectivity (rs-FC). This rs-FC is found to be present among several different brain networks known as resting-state networks, performing important roles  
like vision, language, sensory-motor functions, etc. \cite{beckmann2005investigations}. Some of the resting-state networks (RSN) discovered in previous studies are Default Mode Network (DMN), Frontal Parietal Network (FPN), Cingulo opercular Network (CON), Sensory Motor Network (SMN), Cerebellum Network (CN), and Occipital Network (ON) \cite{biswal1995functional,dosenbach2010prediction}. There have been a lot of aging-based studies on the within and between functional connectivity of these RSNs. Some studies investigated that RSN like DMN, SMN, and CON show decreased within connectivity with aging and increased between connectivity which reflects decreased segregation \cite{chan2014decreased ,geerligs2015brain} .Studies suggest that previous research have focussed on specific age groups in spite of various brain developmental age groups like young, middle young, middle late and elder age group who are more prone to neurodegenerative disorders. Also, generalizing aging-based brain FC on the basis of one specific RSN like DMN is not a good way to understand brain functional organization \cite{greicius2004default ,tsvetanov2016extrinsic} , instead, we need a study based on whole-brain FC networks. It is interesting to note that two brain regions can be positively correlated or anti-correlated, but most of the studies have estimated whole-brain connectivity on the basis of positive correlations, ignoring anti-correlation between brain networks and that might affect the study of segregation and dedifferentiation of various brain networks concerning different brain developmental age groups \cite{chan2014decreased ,tomasi2012aging} .In this work, the resting state of different brain regions has been investigated using a brain template named Dosenbach’s template \cite{dosenbach2010prediction}. In Dosenbach’s study, after a series of meta-analyses on task-based studies, 160 regions of interest) were found which were further distributed into six brain networks using a mechanism of modularity optimization on the average Functional Connectivity matrix of various cohorts of healthy subjects. Similarly, in this study, six brain networks have been discovered using various k-means clustering mechanisms, an unsupervised learning algorithm that can cluster different ROI’s into brain networks, and has been used in the fMRI studies for a long \cite{fan2015functional}. Later t-SNE have been applied to those clustered data to check the clear separation of these clusters \cite{wattenberg2016use}. Meanwhile, various brain developmental stages were discovered using Random forest classification of sample entropy feature calculated for each of the 160 ROI of every individual subject of different age groups. Later, within network connectivity, between network connectivity, and segregation of brain networks are studied for different brain developmental stages using positive and anti-correlation temporal data of different age groups.
The paper has been organized in the following sequence: The Materials and Methods have been presented in section II. Results, conclusion, and discussion in sections III, and IV respectively. 

\maketitle

\section{Materials and Methods}

\subsection{Dataset Description}
Resting state-fMRI data for this study have been acquired from the Human Connectome Project (HCP)  publicly available at http://www.humanconnectomeproject.org/data/.The major aim of HCP is to collect data from a large cohort of subjects ranging in different age groups and further generate an in vivo mapping of functional connectivity to be used in the research domain. The dataset used comprises rs-fMRI data of 1096 healthy subjects lying in the age group of 7-89 years and has been acquired using a 3.0 Tesla Magnetom Tim Trio scanner. The scanning parameters were as follows: Repetition time (TR) = 2530 ms, Echo Time (TE) = 30 ms, Field of View (FOV) = 240 * 240 mm2, acquisition matrix = 64*64, voxel size = 3mm*3mm*4mm, and 34 slices covering the entire brain and the acquisition sequence lasted for 500 seconds per subject. Participants were asked not to fall asleep during the process to collect 3-dimensional functional images of every participant. Lastly, the samples were retained after applying criteria for head motion and other corrections for individual subjects.
After the application of all the required pre-processing steps, the BOLD activation time-series signal of each voxel was extracted from the functional images of the brain \cite{di2013calibrating}. Based on Dosenbach’s Template \cite{dosenbach2010prediction}, one hundred and sixty regions of interest (ROIs) were then priorly selected and the meantime series within the spherical range of 5mm radius was calculated. Finally, the time series data matrix was computed for each subject consisting of 160 brain node data of different timestamps that have been processed further in the study.


\begin{figure*}[t]
\centering
\includegraphics[width =0.8\linewidth]{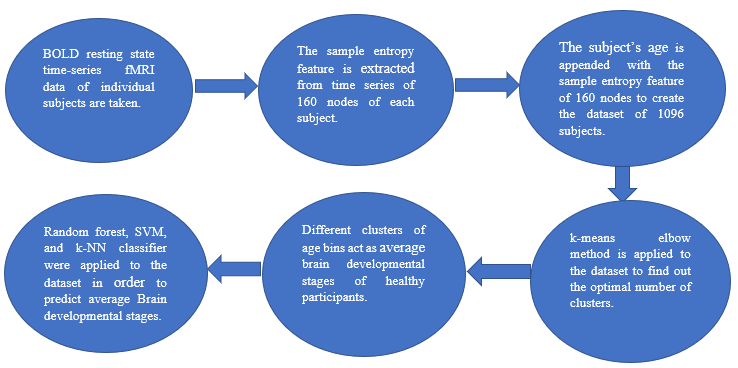}
\caption{Flow chart for brain average developmental stages prediction using Sample entropy.}
\label{figure:block diagram1}
\end{figure*}
\subsection{Statistical Feature extraction}

Figure 1, depicts the flowchart to classify average brain developmental stages of human participants, a new dataset is created using the sample entropy feature extracted from the BOLD activation time series of the 160 dosenbasch ROIs.
Sample entropy ($SampEn$) has been used more often in previous research to investigate signal complexity and irregularity of physiological time series signals because of its length independence property \cite{sokunbi2014sample}.  A time-series data with less amount of repetition or less complexity has a smaller value of $SampEn$ as compared to complex time series data.

Mathematically Sample entropy of a time-series of length N (x1, x2, x3…,xN) is computed using given set of equations, 
 \begin{equation*}
  SampEn(m,r,N)= -\ln [\frac{(U^{(m+1)} (r))}{(U^m  (r) }]
   \end{equation*}
  \begin{equation}
    U^m  (r)= [N-m\tau]^{-1 } \sum_{i=1}^{N-m\tau} C^m_i (r)
\end{equation}\\

 Where,
\begin{equation}
   C^m_i (r) = \frac{B_i}{N-(m+1)\tau}
\end{equation}\\
\begin{center}
  $ B_i = $ number of j where,
\begin{equation}
  d|Xi,Xj|\leq r  
\end{equation}
                          
 $X_i  = (x_i,x_{i+\tau}  ,…,x_{i+(m-1)\tau})$\\
 $X_j  = (x_j,x_{j+\tau}  ,…,x_{j+(m-1)\tau})$\\
 $1\leq j \leq N-m\tau,j \neq i$\\  
\end{center}

 \vspace{2mm}

 Here, N specifies the data length, m denotes pattern length, r and $\tau$  are tolerance value and time delay respectively and $X_i$ and $X_j$ are pattern vectors. Two patterns i and j of m measurements of the time series are similar if the difference, $ d|X_i,X_j| \leq r$.

After extracting SampEn of 160 nodes of 1096 subjects, age feature is appended to form the required dataset and following that k-means elbow criterion \cite{kodinariya2013review} defined as the ratio of within-cluster distances to between cluster distances is applied to find the optimal number of clusters of brain developmental stages as shown in Figure 2(a), indicating elbow at k = 4 ( with a distortion score of 28790.217) that means four average brain developmental stages with following age bins: (7-19) years, (20-34) years, (35-53) years, and (54-89) years respectively. Figure 2(b), depicts the two-dimensional scatter plot of 4 clusters obtained using $SampEn$ data of participants. 

\subsection{Prediction of Brain developmental stages }
Neurodegenerative disorders may lead to abnormal development of the human brain that usually yields abnormal brain signals and it’s more prominent in the $4^{th}$ brain developmental stage where the age group lies beyond 60 years. So, if we can predict the abnormality in the brain developmental stage of any participant (in the initial stage of any such neurological disorder) by comparing with same age group healthy subject’s brain developmental stage, this can be of immense help. For that purpose, we have classified the rs-fMRI entropy dataset appended with age into four brain developmental stages using different classifiers and we can predict the four stages with good accuracy as presented in figure 2(c). Before classification, some pre-processing steps dimensionality reduction using PCA was performed, considering only such dimensions containing 90 percent of the total variance in the data. Further, after dimensionality reduction, the dataset is split into training and testing datasets and then classified with Random Forest classifier, SVM (Support Vector Machine), and k-NN (k-Nearest Neighbourhood). Since the dataset used was skewed as different developmental stages age bins have an uneven number of participants, for example, Class1 - 276, Class2 - 647, Class3 - 118, and Class4 – 57, and SVM supports the classification of the skewed dataset by minimizing the cost of misclassification and assigning different weights to classes, so, we started the classification with SVM, but again, since it is a multi-class problem and  Random forest is intrinsically suited for that purpose, so, it gave better accuracy than other classifiers.In this way, we were able to predict the brain developmental stage of a subject (with an accuracy of 91.7 percent) based on its rs-fMRI data which might help to draw some inferences about the growth of the disorder at an early stage.

\begin{figure*}[t]
\centering
\begin{subfigure}[h]{0.5\linewidth}
\includegraphics[width =\linewidth]{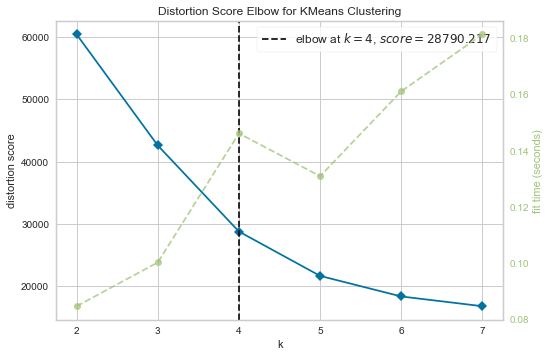}
\label{figure:blockd2}
\caption{K-means elbow clustering method yielding 4 optimal clusters. The X-axis denotes the range of clusters(k), the y-axis denotes the distortion score.}
\end{subfigure}
\begin{subfigure}[h]{0.45\linewidth}
\includegraphics[width =\linewidth]{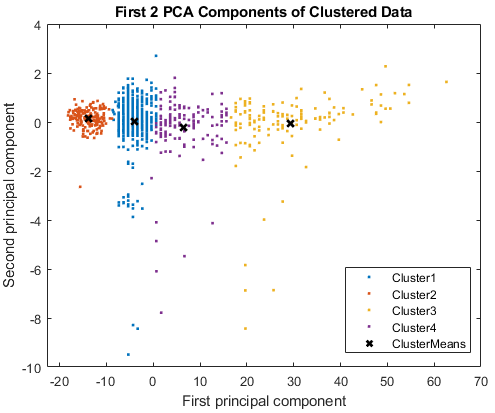}
\label{figure:figure 3}
\caption{2-D scatter plot of k-means clustered data. } 
\end{subfigure}
\begin{subfigure}[h]{0.6\linewidth}
\includegraphics[width =\linewidth]{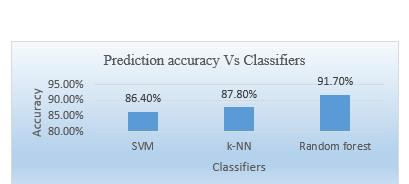}
\label{fig:figure 4}
\caption{Brain developmental stages prediction accuracy of different classifiers.} 
\end{subfigure}
\label{figure 2}
\caption{Clustering, Plotting and Prediction of brain developmental stages using $SampEn$ values.}
\end{figure*}
\begin{figure*}[t]
\centering
\includegraphics[width =0.9\linewidth]{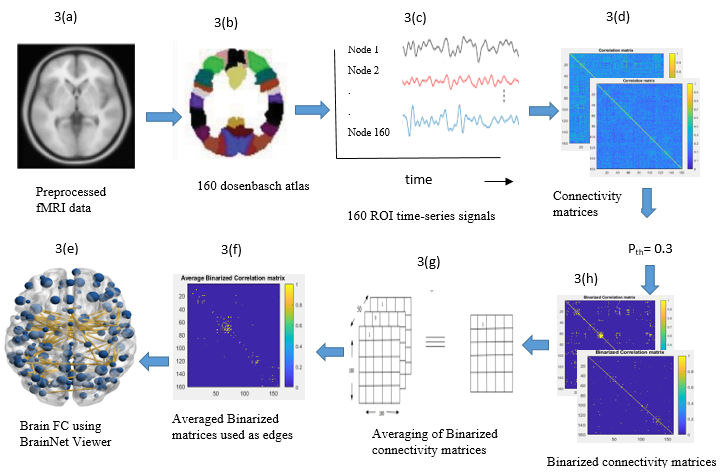}
\caption{Flow chart of brain network visualization}
\label{figure:figure 5}
\end{figure*}
  
\subsection{Brain Network visualization of developmental stages}
To visualize the average brain networks of each age cohort with an aim of ensuring almost equivalent and a proportional number of subjects in each age group as per the HCP dataset, we have done the following categorical divisions: Younger adults (YA): (7-19 years, n= 63), Middle Adult (MA): (20-34 years, n=75), Middle  
Middle late (ML) :(35-53 years, n=70), and Elderly (E): (54-89 years, n = 40). We have used a graph theoretical network visualization toolbox, BrainNet Viewer (https://www.nitrc.org/projects/bnv/) to illustrate human connectomes as ball and stick models \cite{xia2013brainnet}. The following methodology has been adopted to visualize the brain networks at 4 developmental stages: firstly, Raw rs-fMRI time-series data of every participant based on Dosenbasch brain anatomical parcellation is taken. Then, the Pearson correlation between different ROI time series data is computed. After that, threshold analysis based on a graphical approach (number of edges versus threshold value) is done and an absolute threshold (0.3) is found, although, in the past,  researchers have described different thresholding techniques \cite{azarmi2019granger}  and have used absolute threshold value such as 0.2 for binarizing the correlation matrix \cite{xia2013brainnet}. Following that, the Binarized correlation matrix for different subjects of each age cohort is averaged and an edge file is created which is further used in the brain network visualization as shown in figure 3. Figure 4 shows the brain network connectivity of different brain developmental stages and it is clearly depicted that the brain connectivity follows the decreasing order of $ MA>ML>YA>E$. To understand these connectivity changes across the lifespan, we need to understand different brain networks’ connectivity individually.

\begin{figure*}[t]
\centering
\begin{subfigure}[h]{0.23\linewidth}
\includegraphics[width =\linewidth, keepaspectratio = True]{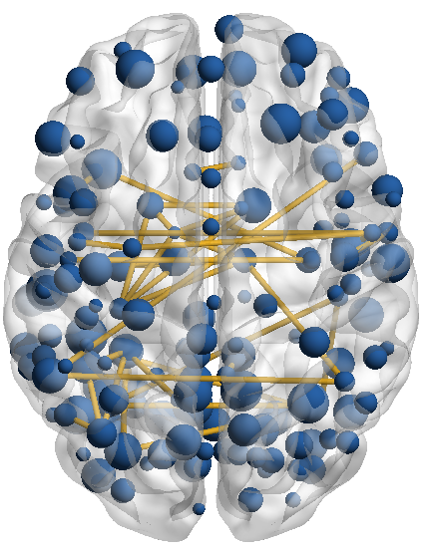}
\label{figure:figure 6a}
\caption{Young Adults (7-19 years)}
\end{subfigure}
\begin{subfigure}[h]{0.23\linewidth}
\includegraphics[width =\linewidth, keepaspectratio = True]{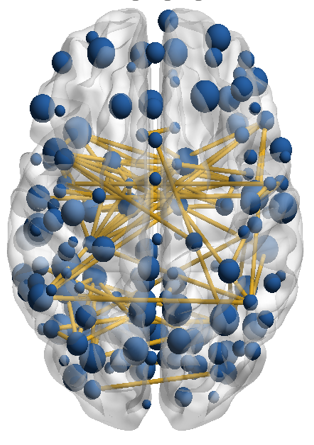}
\label{figure:figure 6b}
\caption{Middle young (20-34 years)} 
\end{subfigure}
\begin{subfigure}[h]{0.23\linewidth}
\includegraphics[width =\linewidth, keepaspectratio = True]{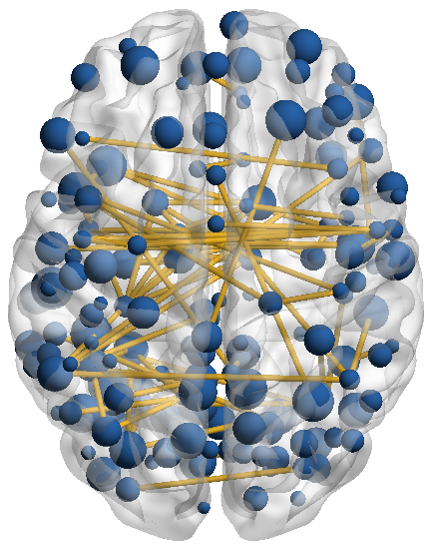}
\label{fig:figure 6c}
\caption{Middle late (35-54 years)} 
\end{subfigure}
\begin{subfigure}[h]{0.23\linewidth}
\includegraphics[width =\linewidth, keepaspectratio = True]{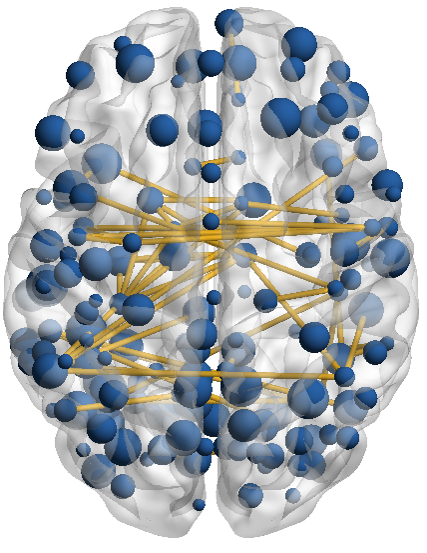}
\label{fig:figure 6d}
\caption{ Elder (55-89 years)} 
\end{subfigure}
\label{figure 4}
\caption{ (a),(b), (c) and (d) represents brain network connectivity for 160 dos ROI's for YA, MA, ML, and E stages respectively.}
\end{figure*}

\begin{figure*}[t]
\centering
\begin{subfigure}[h]{0.45\linewidth}
\includegraphics[width =\linewidth, keepaspectratio = True]{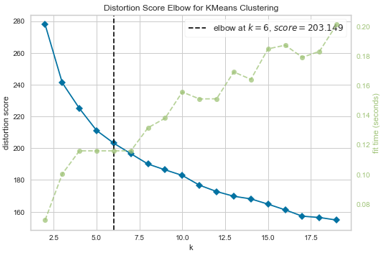}
\label{figure:figure 7}
\caption{ K-means elbow applied on data matrix to obtain optimum six clusters.}
\end{subfigure}
\begin{subfigure}[h]{0.4\linewidth}
\includegraphics[width =\linewidth, keepaspectratio = True]{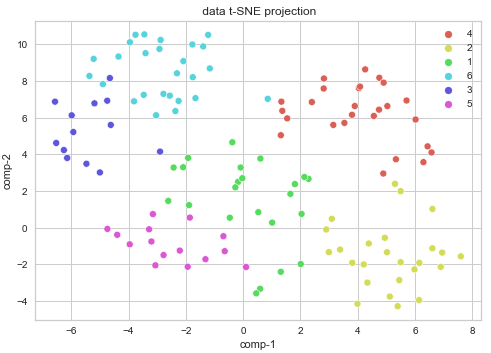}
\label{figure:figure 8}
\caption{ Data t-SNE projection of clustered data forming six clusters} 
\end{subfigure}
\label{figure 5}
\caption{Optimum clusters of 160 dos ROI data matrix and visual separation of those clusters using t-SNE method. }
\end{figure*}

\begin{figure*}[t]
\centering
\includegraphics[scale=0.9, keepaspectratio=True]{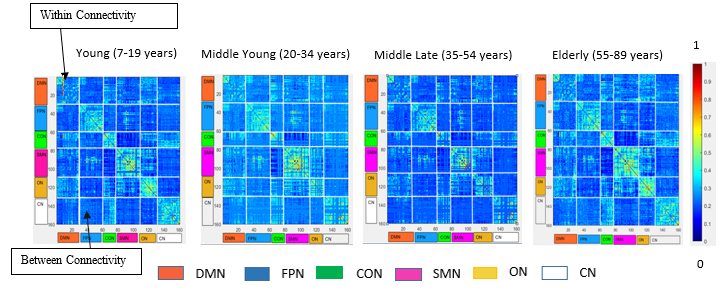}
\caption{ Mean node to node adjacent data matrix (mean connectivity, r(z)) based on six clusters for each brain developmental stage. Six clusters have been labelled using colorbars along axes as DMN (Default mode Network), FPN (Frontal Parietal Network), CON (Cingulo Opercular Network), SMN (Sensory Motor Network), ON (Occipital Network), and CN (Cerebellum Network) based on\cite{dosenbach2010prediction}}
\label{figure:figure 6}
\end{figure*}

\begin{figure*}[t]
\centering
\includegraphics[width =0.9\linewidth]{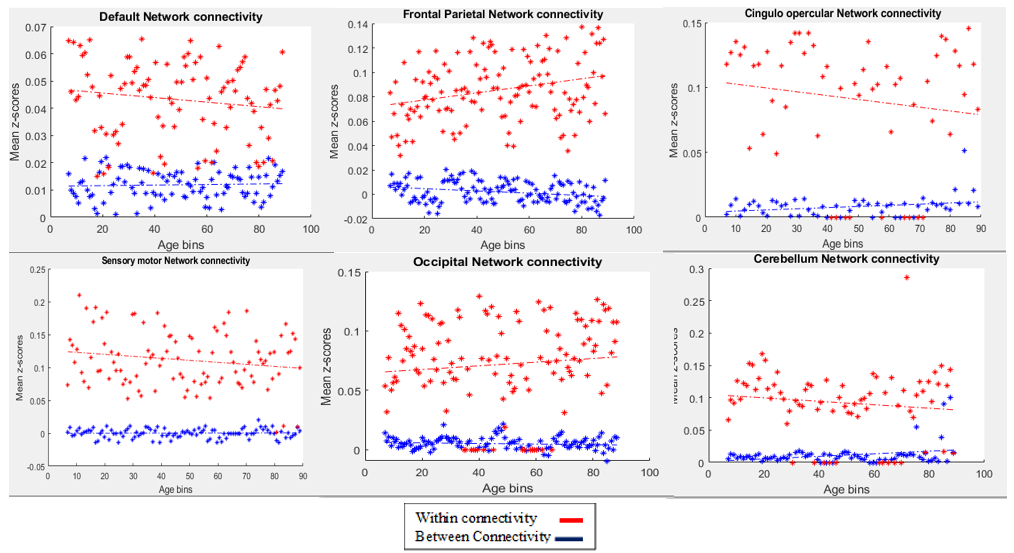}
\label{figure:figure 7}
\caption{  Mean System Connectivity with respect to age for six dosenbasch brain networks, DMN, FPN, CON, SMN, ON, and CN respectively. X- axis: Age (years), y-axis (Mean z-score).}
\end{figure*}
\begin{figure*}[t]
\centering
\includegraphics[width =0.9\linewidth]{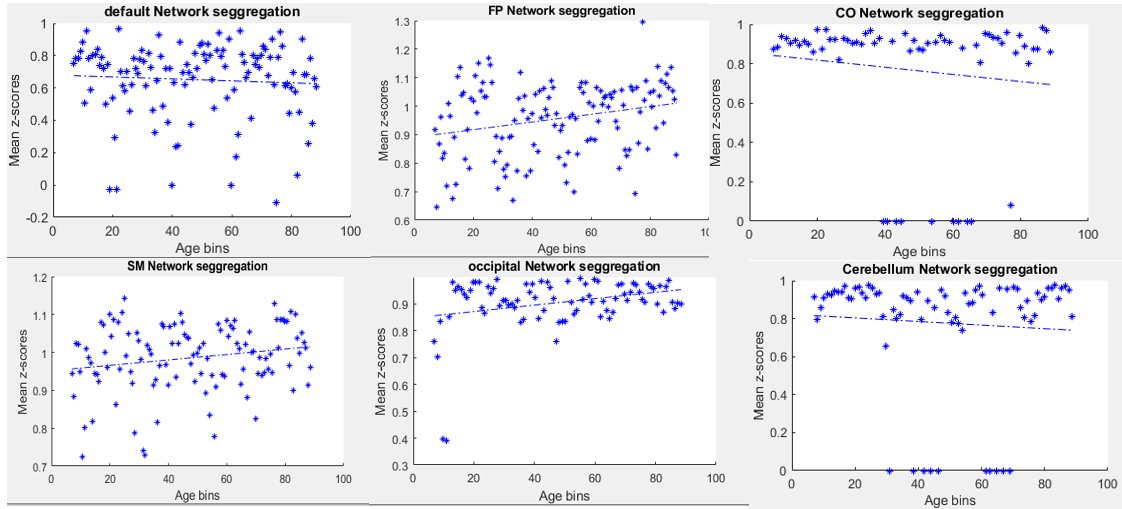}
\label{figure:figure 8}
\caption{ Mean Network Segregation with respect to age for six dosenbasch brain networks, DMN, FPN, CON, SMN, ON, and, CN respectively. X- axis: Age (years), y-axis (Mean z-score).}
\end{figure*}

\subsection{Clustering ROIs into Six Functional Networks}
For every participant of each brain developmental stage, there exists a static functional connectivity matrix of size 160*160, where the static FC between each pair of ROIs (nodes) is computed using Pearson’s correlation coefficient between time-series data of two ROI’s. These correlation coefficients were converted into Fisher’s z-transformed r-matrix using Fisher’s equation \cite{zar1996confidence}. The resulting r-matrix is a fully connected matrix forming a weighted relatedness graph. In the past, researchers have excluded negative correlations \cite{chan2014decreased} but since those values hold significance in whole-brain FC \cite{rubinov2011weight}, it has been included in the present study for the clustering purpose. The final data matrix for each subject forms a 160 × 160 z-matrix with the diagonal values set to zero and is used further for clustering 160 ROIs into an optimal number of clusters. K-means elbow algorithm is applied to the dataset of individual participants to obtain an optimum of six clusters as shown in figure 5(a). Based on these clusters, nodes were reordered to form an adjacent matrix for every participant. Since, the study includes a different number of subjects in each brain developmental stage, a group or average adjacent matrix for each stage is formed and the K-means algorithm is applied further to the group adjacent matrix to obtain group clusters whose separation is validated and visualized using t-SNE (t-distributed stochastic neighbourhood embedding) algorithm in figure 5(b). t-SNE\cite{wattenberg2016use} is applied on the adjacent group matrix with default tuning hyper parameters of initial dimensions (50), perplexity (30), maximum iterations (1000), and dimensionality reduction have been performed to convert high dimensional data into two dimensions forming two principal components that have been visualized using a scatter plot showing significant separation of clusters. Six clusters of group adjacent matrix (reordered 160 nodes) form the six functional resting-state brain networks for each age cohort as shown in figure 6. 

\section{Results}
\subsection{Within system Connectivity across different brain developmental stages on a network level}
Correlation values across the diagonal of the mean adjacent matrix represent within system functional connectivity. For different clusters or resting-state brain networks, within system connectivity varies differently across different age cohorts and are not homogeneous. As we can see, in figure 6, Within system connectivity across DMN, SMN and CON decreases with age whereas , it increases for ON and FPN and similar results have been reported in previous studies using different Atlases \cite{zonneveld2019patterns}. For network like CN, it remained almost constant for different brain developmental stages. In order to find out the pattern of how brain networks within connectivity change with age, mean node to node z- values of all nodes of a system to each other were computed. linear fits were applied to the within connectivity data after removing outliers using the percentile detection method and it was found that the age function for these networks was fitting significantly using a linear model as shown in figure 7.

\subsection{Between system Connectivity across different brain developmental stages on a network level}
Correlation values across off-diagonal of the mean adjacent matrix represent between-system functional connectivity. Similar to within system connectivity, between system connectivity also varies for different brain networks as the brain developmental stage changes as shown in figure 6, but in most of the networks, this connectivity increases to some extent, and for some networks, it remains almost same. In order to find out the pattern of how brain networks between connectivity change with age ,between-system connectivity was calculated as the mean node to node fisher's z-value between each node of a system and all the nodes of rest other systems. After that, linear fits were applied to the between system connectivity data \cite{chan2014decreased,zonneveld2019patterns} after removing outliers using the percentile detection method and defined upper and lower thresholds and it was found that the age function for these networks was fitting significantly using a linear model as shown   in figure 7. 

\subsection{Network Segregation for different brain developmental stages}
To study the correlation between within-network connectivity (WNC) and between network connectivity (BNC) as the age progresses, segregation can be used as a network measure (Chan et al., 2014). Mean network segregation can be described as the ratio of the difference in mean within network correlations (z-value) and between network correlations with mean within network correlation values, as given below: 
\begin{equation}
   Network Segregation (NS)=\frac{(\Bar{WNC}-\Bar{BNC})}{\Bar{WNC}}                 
\end{equation}

If the NS value is high it reflects a lower correlation between WNC and BNC for a specified network, similarly, if this NS value is low, it indicates a stronger correlation value between WNC and BNC. Also, NS values less than 0, indicate a higher correlation between WNC and BNC i.e., diminished segregation of networks.

In order to see the effect of aging or changes in brain linear fits were applied to the NS values on a network level, after removing the outliers using percentile detection method in MATLAB and it was found that older age is associated with decreased segregation for most of the networks except few as shown in figure 8. This is because of the increase in the integration of the networks with respect to aging. So, we can say that, age related decrease in the connectivity of some brain networks are accompanied by an increase the in the connectivity of other resting-state brain networks, in a nutshell, overall rs- functional connectivity reorganizes with changes in brain developmental stages or shows a compensatory mechanism with aging \cite{sala2015reorganization , song2014age} . 

\section{Conclusion and Discussion}
  The findings of the present study on rs-fMRI time series data indicate that the brain is a complex network sub-divided into various clusters or rs-Networks and the K-Means clustering algorithm is a feasible method to find out those clusters and four brain developmental stages of the human brain. Brain developmental stages can be best predicted using sample entropy appended with age as a feature by Random forest classifier (accuracy, 91.7 percent). Also, these average brain developmental stages are accompanied by changes in brain network connectivity and network segregation in such a way that increases and decreases in connectivity follow a compensatory mechanism. Graph theory-based BrainNet viewer can be used significantly to study the changes in the number of connections between brain regions as age progresses. Specifically, on a network-based analysis, the Default mode network within connectivity decreases with age but between network connectivity increases to some extent, conversely, the Occipital network shows an increase in within network connectivity. So, we found a pattern of increase and decrease of connectivity within and between networks in equal proportion \cite{zonneveld2019patterns}. Our study can add an understanding of changes in functional connectivity of an aging brain as the brain developmental stage changes from middle-late to the elderly group and can serve as a strong basis for studies investigating the role of FC as a potential early marker for neurodegenerative diseases like Dementia, Alzheimer, etc. This study has some limitations that need to be addressed, we have used very few network measures to study the effect of change in brain developmental stages on brain network connectivity, in the future, some other brain network measures will be added to get some more significant results. Also, the study is purely on the basis of resting-state fMRI data, we can add some task-based fMRI data to study the generalized effect of aging on human brain signals. Along with that changes in structural connectivity of the brain with changes in brain developmental stages can be studied and combined with previous results to see the exact anatomical changes with the aging

\bibliographystyle{IEEEtran}
\bibliography{refs.bib}

\end{document}